\documentclass[twocolumn]{aastex61}

\usepackage{color}
\usepackage{bm} 
\usepackage{amsmath,amssymb}

\begin{document}

\title{Photospheric response to a flare}

\author{Michael S. Wheatland}
\affiliation{Sydney Institute for Astronomy\\  School of Physics\\University of Sydney 
\\NSW 2006 \\Australia}

\author{Don Melrose}
\affiliation{Sydney Institute for Astronomy\\  School of Physics\\University of Sydney 
\\NSW 2006 \\Australia}

\author{Alpha Mastrano}
\affiliation{Sydney Institute for Astronomy\\  School of Physics\\University of Sydney 
\\NSW 2006 \\Australia}

\newcommand{\red}[1]{{\color{red}{\textit{#1}}}}

\begin{abstract}
Flares produce sudden and permanent changes in the horizontal photospheric magnetic field. In particular flares generally produce increased magnetic shear in the photospheric field along the neutral line. Recent observations show also that flares can produce sudden photospheric motion. We present a model for the observed changes as the response of the photosphere to a large-amplitude shear Alfv\'{e}n wave propagating down from the corona on either side of the neutral line. The Alfv\'{e}nic front is assumed to impact the photosphere close to the neutral line first, and then successively further away with time, such that the line of impact coincides with the flare ribbon. The wave introduces magnetic shear and velocity shear. The magnetic shear introduced at the photosphere has the same sign on either side of the neutral line, while the velocity shear has the opposite sign. We discuss the possibility that this process is responsible for particle acceleration in flares.\end{abstract}

\section{Introduction}\label{sec:introduction}

During the impulsive phase of a solar flare, magnetic energy is converted into other forms of energy in the solar corona. The accepted mechanism underlying flares is magnetic reconnection, a process in which coronal magnetic field lines change their connectivity. Despite decades of investigation, many details of the flare process remain poorly understood~\citep{2017LRSP...14....2B}.

It is difficult to measure the coronal magnetic field directly, but photospheric magnetic field measurements have revealed that flares produce sudden and permanent changes in the observed magnetic field~\citep[e.g.][]{2005ApJ...635..647S}. The most detailed information comes from vector magnetogram measurements, which show that the predominant change in a flare occurs in the horizontal magnetic field, which tends to increase parallel to the neutral line, i.e.\ the magnetic shear along the neutral line increases~\citep{2012ApJ...745L..17W,2012ApJ...759...50P}. There are corresponding sudden changes in the electric current density close to the neutral line~\citep[e.g.][]{2016A&A...591A.141J}.

The observations have been interpreted as the photospheric response to coronal magnetic restructuring during the flare. Changes in the photospheric field values imply changes in the net Lorentz force on the corona (which can be calculated from the boundary values of the field), and the values of the changes have been used to interpret the effect of the flare in the corona~\citep{2012SoPh..277...59F,2016SoPh..291..791P,2016ApJ...820L..21X}.

\citet{2016NatCo...713104L} reported a striking example of changes at the photosphere during the 22 June 2016 M6.5 flare: a sunspot was observed to rotate suddenly in response to the passage of a flare ribbon across the spot. The observations confirm that coronal field changes can produce not only photospheric field changes, but also substantial induced motion of the dense photosphere, contrary to general expectations~\citep{2016NatPh..12..998A}.
Other examples of flare-induced sunspot rotation have also been reported, including the case of a sunspot reversing its direction of rotation~\citep{2016NatCo...713798B}. 

Hard X-ray (HXR) observations of flares imply that a significant fraction of the released energy goes into accelerated electrons with energies $10-100$\,keV~\citep{2017LRSP...14....2B}. It is generally assumed that the electrons originate high in the solar corona, perhaps at the site of magnetic reconnection, and then follow field lines down to the dense lower atmosphere, where they produce hard X-rays via thick-target bremsstrahlung~\citep{1971SoPh...18..489B}. However, this picture for HXR production suffers from the ``number problem.'' Because the electrons originate in the low-density corona, the implied particle fluxes at the low atmosphere would evacuate electrons from a substantial volume above an active region during a flare~\citep{1976RSPTA.281..473B}. A return current of electrons from the dense chromosphere to the corona is required, but this also introduces problems. The observations of flare-induced photospheric motion imply that energy is also transported from the corona to the photosphere in other forms. The photospheric changes occur behind the flare ribbons, the site of hard X-ray emission, which suggests a more direct connection between the magnetic field change at the photosphere and the acceleration process.

In this article we present a simple 2D magneto-hydrodynamic model for the response of the photosphere to a flare, in terms of a large-amplitude shear Alfv\'{e}n wave produced by coronal reconfiguration impacting the photosphere, and introducing a magnetic and velocity shear close to the neutral line. To motivate the model we return to the observations of the 22 June 2015 event (Section~\ref{sec:analysis}). We present a summary analysis of the observations, as well as the results of nonlinear force-free modeling. In Section~\ref{sec:model} we give the details of the model, and in Section~\ref{sec:discussion} we discuss the model results, and a possible connection to electron acceleration in flares. In Section~\ref{sec:conclusions} we draw conclusions.


\section{22 June 2015 Flare}\label{sec:analysis}

On 22 June 2015 an M6.5 flare occurred in NOAA AR 12371 (event SOL2015-06-22T18:23), accompanied by an eruption and a halo CME. A description of various observations of the flare, and an interpretation of events in terms of a reconnection model, is given in~\citet{2017ApJ...842L..18J}.

\citet{2016NatCo...713104L} presented observations with the high resolution 1.6m Goode Solar Telescope at Big Bear Observatory which show that the 22 June flare caused a sudden rotation of a sunspot to the east of the neutral line involved in the flare. The spot was observed to rotate differentially as the flare ribbon swept across the spot. The rotation was interpreted in terms of a torque exerted on the photosphere by the coronal magnetic field~\citep{2016NatPh..12..998A}.

\begin{figure*}
\gridline{\fig{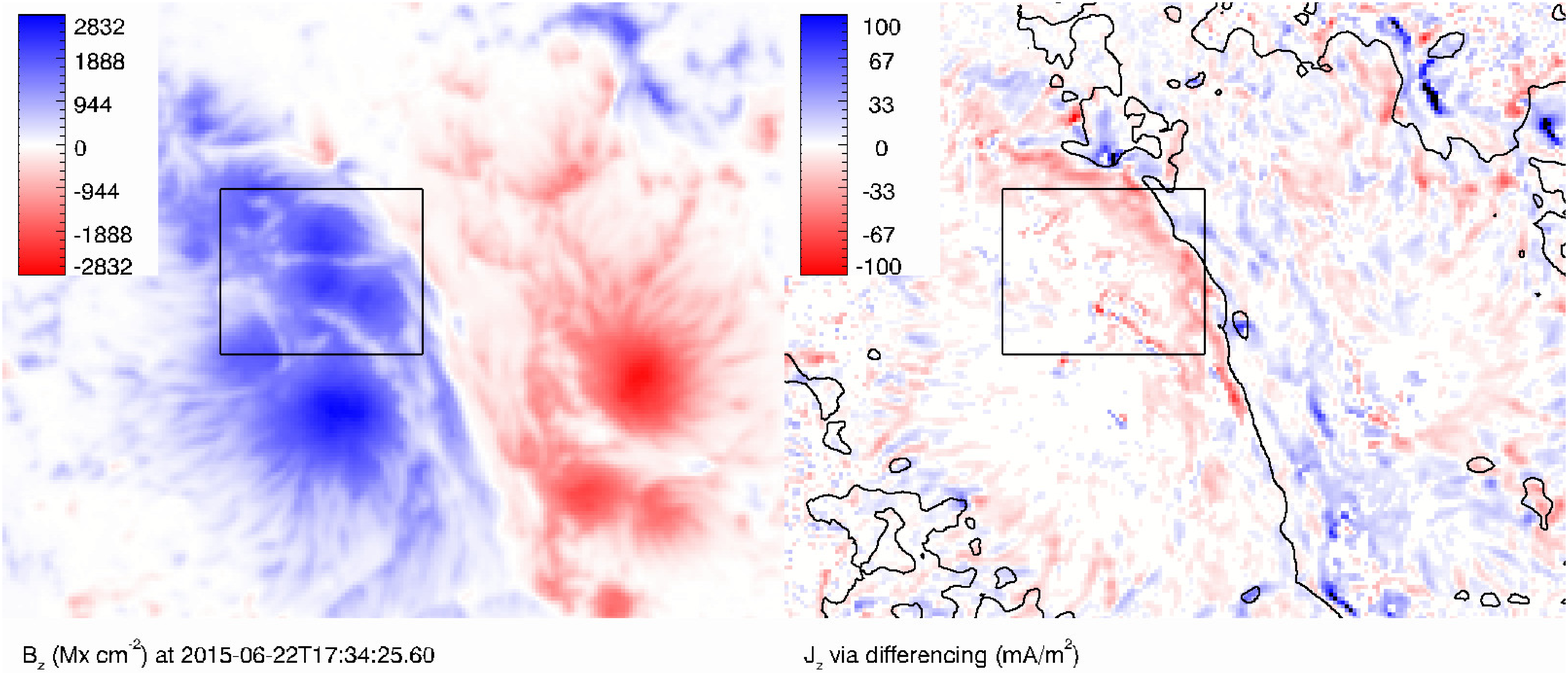}{0.75\textwidth}{(a)}
}
\gridline{\fig{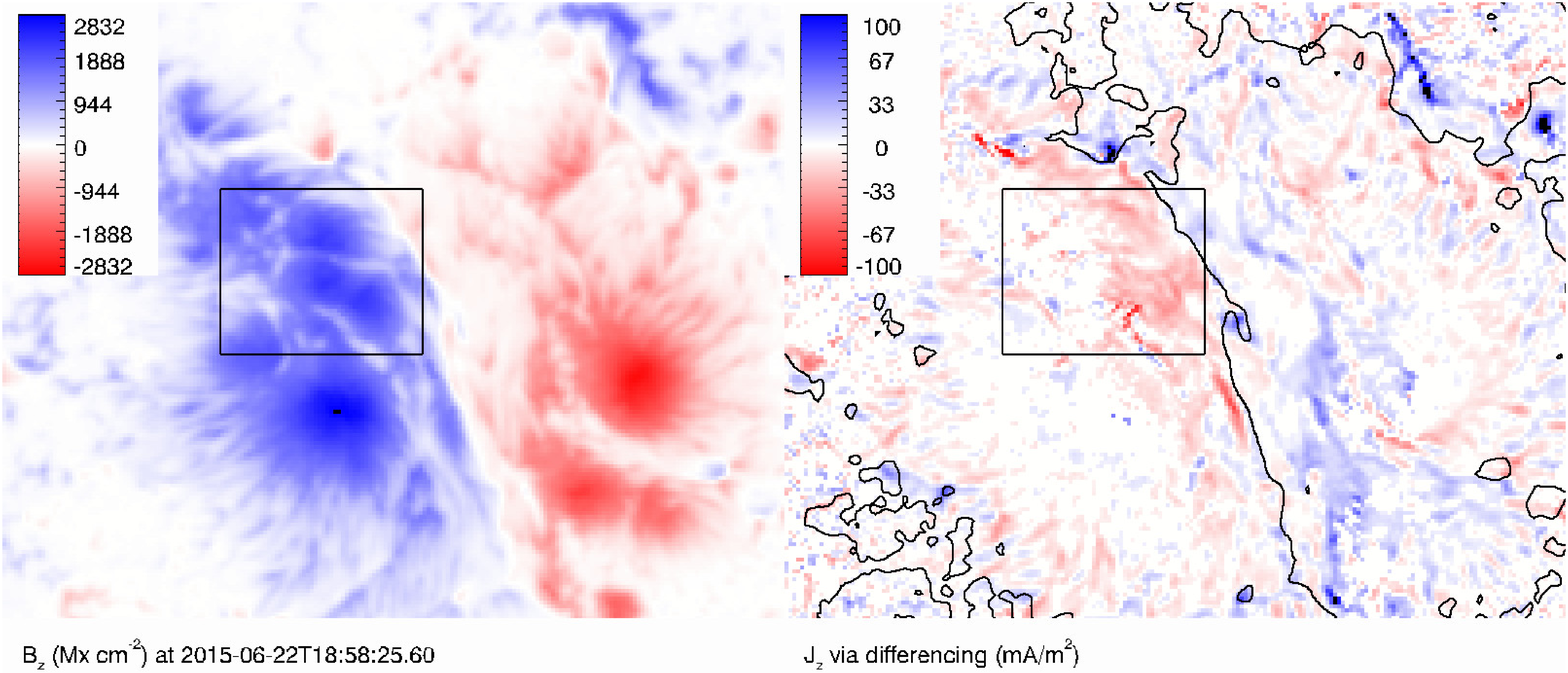}{0.75\textwidth}{(b)}
}
\caption{Vertical component of the magnetic field and vertical component of the electric current density in AR 12371 (a) before and (b) after the 22 June 2015 flare, from SDO/HMI. In each case the left-hand panel shows $B_z$ and the right-hand column shows $J_z$. The box contains the sunspot umbrae described in \citet{2016NatCo...713104L}.\label{fig:f1}}
\end{figure*}

Figure~\ref{fig:f1} shows vector magnetogram data from the Helioseismic and Magnetic Imager on the Solar Dynamics Observatory (SDO/HMI). We use the Spaceweather HMI Active Region Patch (SHARP) data with cylindrical equal area projection (\verb+hmi.sharp_cea_720s+). The top row shows data before the flare (17:34UT) and the bottom row shows data after (18:58UT). The left-hand column shows the locally vertical component of the magnetic field, $B_z$, and the right-hand column shows the vertical component of the electric current density $J_z$ at locations where the signal-to-noise ratio in $J_z$ is greater than one. In the panels showing $J_z$,
the neutral line is indicated by a black solid curve. 

The data in Figure~\ref{fig:f1} show the sudden appearance, coincident with the flare, of a patch of negative electric current density with magnitude $|J_z|\lesssim 50$\,mA/m$^2$ at the location of the rotating sunspot.

\begin{figure*}
\gridline{\fig{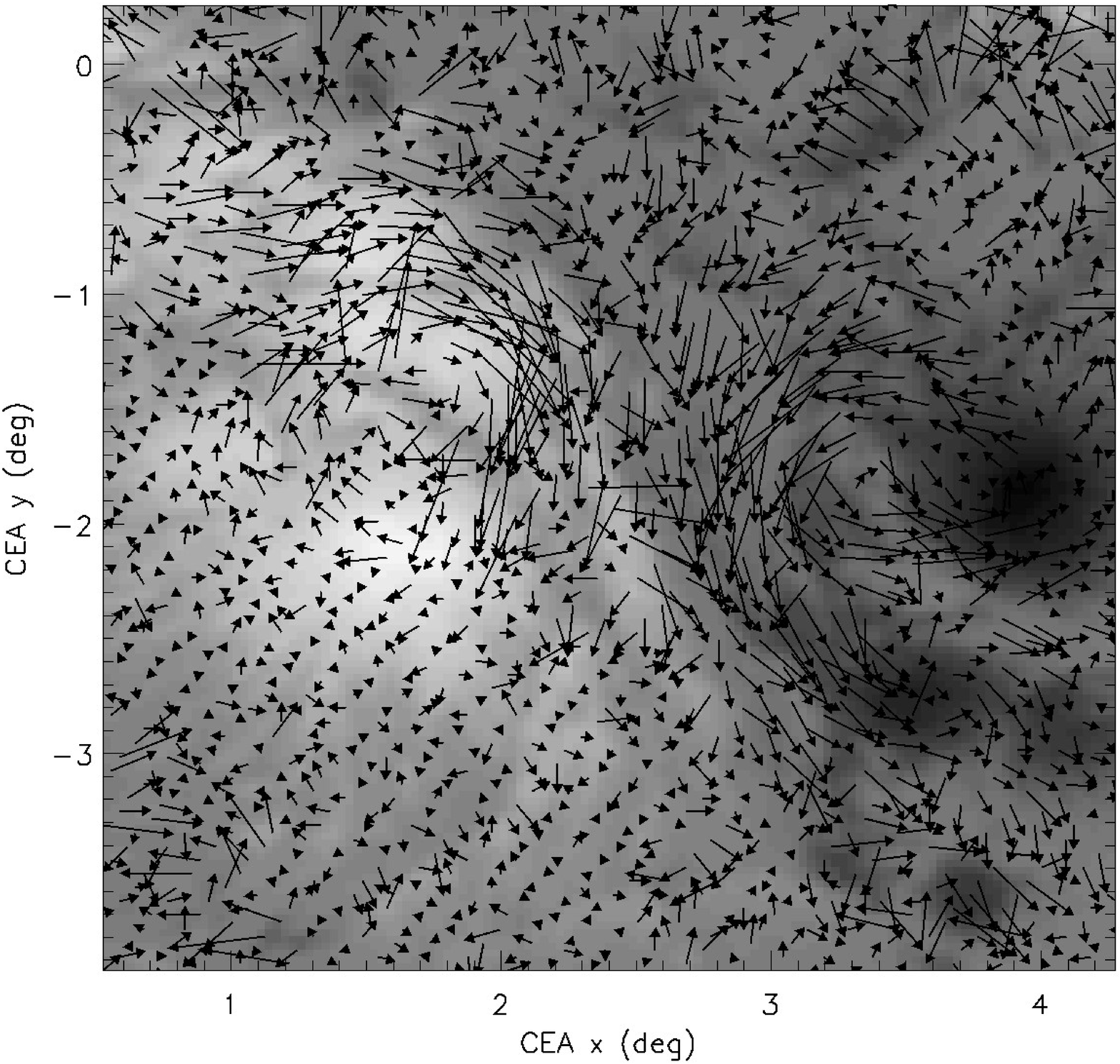}{0.5\textwidth}{(a)}
}
\gridline{\fig{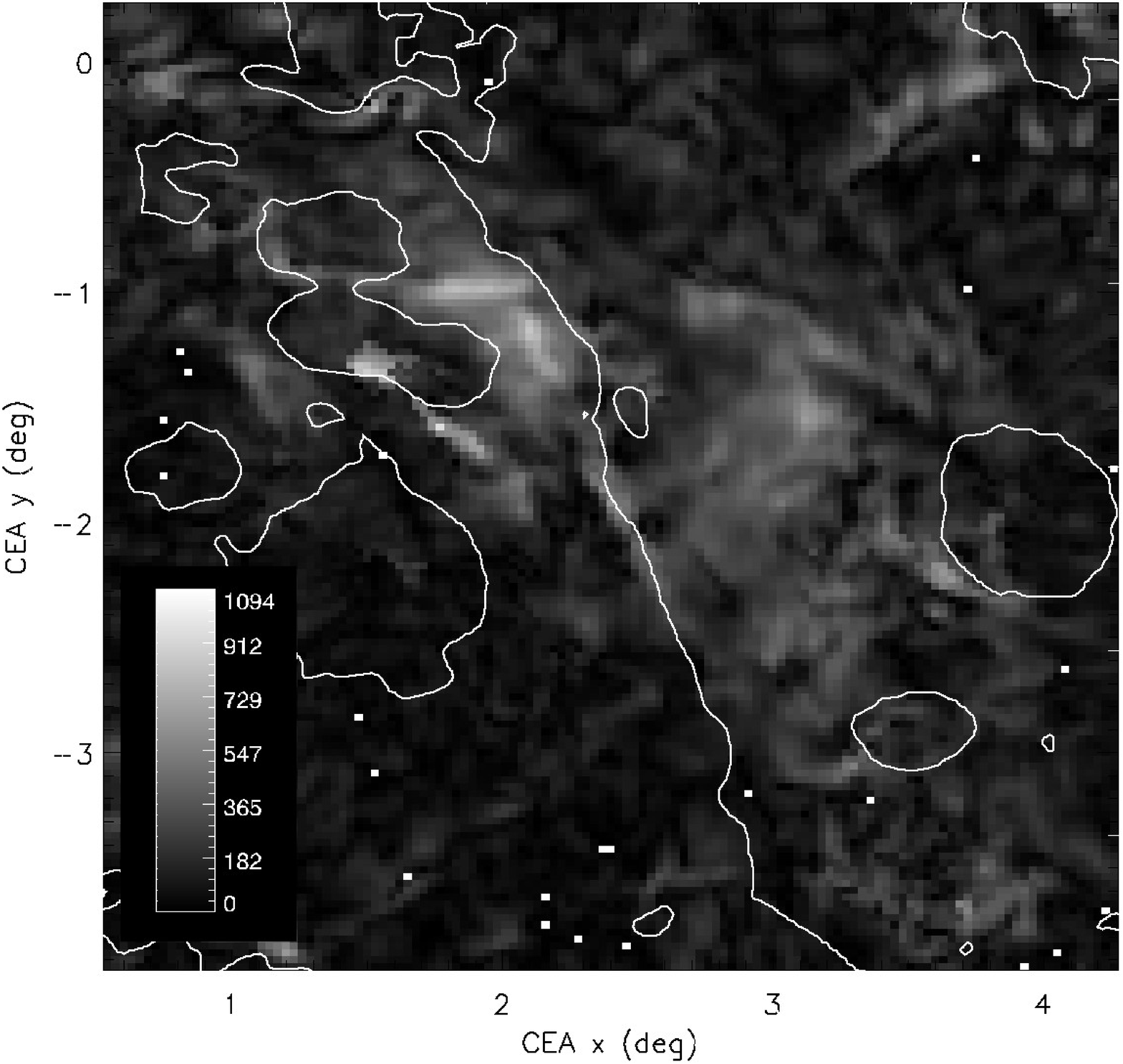}{0.5\textwidth}{(b)}
}
\caption{(a) Vector change in the horizontal field between the two times shown in Figure~\ref{fig:f1}. 
(b) Magnitude of the change.\label{fig:f2}}
\end{figure*}

Figure~\ref{fig:f2} illustrates the corresponding change in the horizontal field $\bm{B}_{\rm h}=(B_x,B_y)$. Panel (a) shows the vector change $\Delta \bm{B}_{\rm h}$ between the two times (before and after the flare) shown in Figure~\ref{fig:f1}. The field of view is a smaller than in  Figure~\ref{fig:f1}, centred on the neutral line. Panel (b) shows the magnitude of the change, $|\Delta \bm{B}_{\rm h}|$. Panel (b) also shows contours of $B_z$ at levels $-1600$, $0$, and $1600$\,gauss, which allow identification of the locations of the large changes in $\bm{B}_{\rm h}$. The flare is seen to introduce a strong shear component along the neutral line, generally directed in the southward direction. The shear is particularly strong close to the neutral line near the sunspot penumbra which rotates. The maximum change in the horizontal field is $\approx 1000$\,gauss. 

Figures similar to panel (b) of Figure~\ref{fig:f2} were presented by \citet{2018ApJ...853..143W} based on SDO/HMI full-disk data -- see Figures 1 (d) and (e) of that paper. \citet{2018ApJ...853..143W} also showed (using high resolution images in the TiO band obtained with the Goode Solar Telescope combined with flow tracking), that the increase in the horizontal field was accompanied by oppositely directed shear flows on either side of the neutral line.

The data show that the flare introduces a strong shear component in the horizontal photospheric magnetic field along the neutral line. The photospheric plasma is also set in motion. The increased shear in the field at the photosphere is assumed to be caused by the introduction, due to the flare, of a horizontal field component in the overlying corona, which is then imposed on the photosphere.
 
Using magnetograms obtained in the near infrared, \citet{Xu-etal-2018} also identified, during the 22 June 2015 flare, photospheric locations with transient changes in the azimuthal direction of the horizontal magnetic field. The changes occurred when the flare ribbons propagated across the sites.

\begin{figure}[t!h]

\gridline{\fig{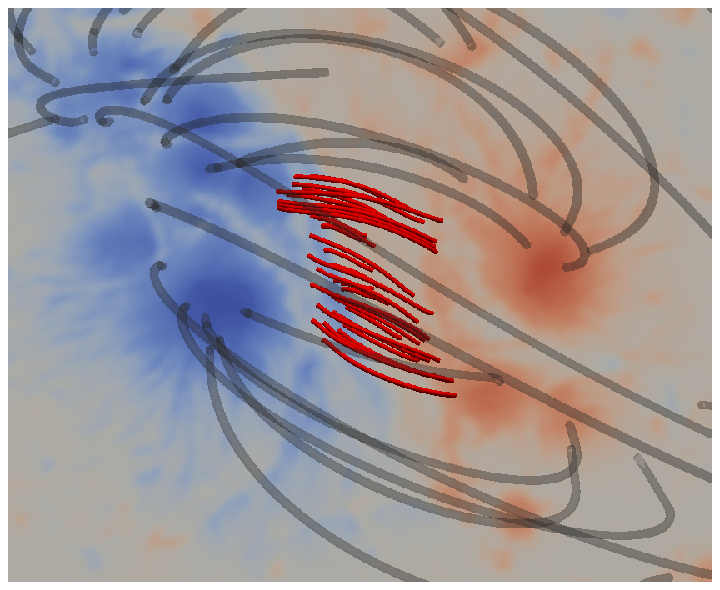}{0.2\textwidth}{(a)}
  \fig{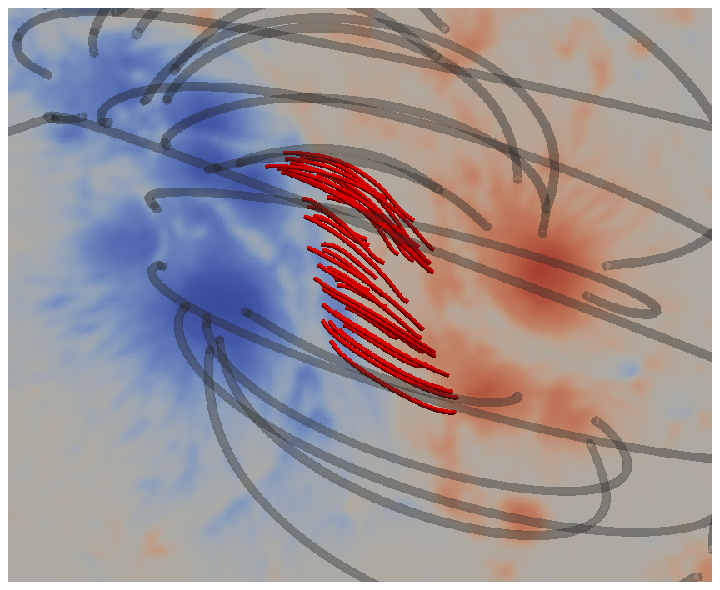}{0.2\textwidth}{(b)}
  }
\gridline{\fig{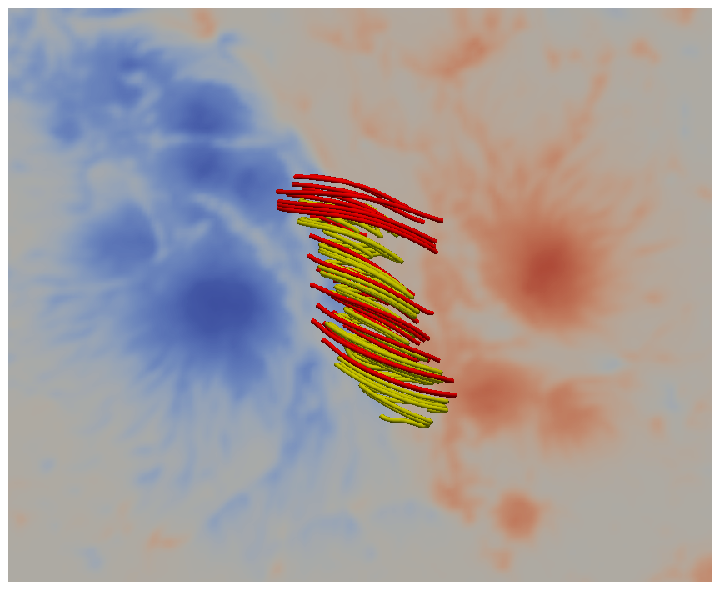}{0.2\textwidth}{(c)}
  \fig{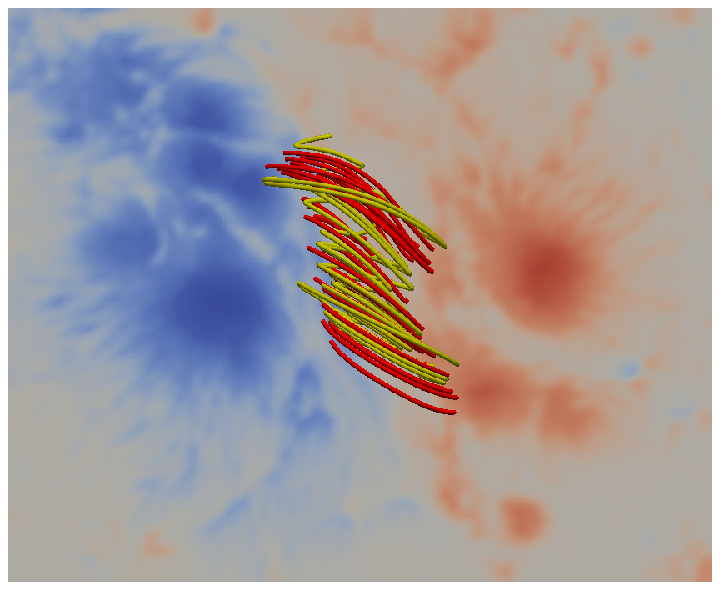}{0.2\textwidth}{(d)}
  }

  
\caption{ Self-consistent NLFFF solutions for the SDO/HMI SHARP data at the two times shown in Figure~\ref{fig:f1}. Panel (a) is before the flare (17:34UT) and panel (b) is after the flare (18:58UT). Two sets of magnetic field lines are shown: a set with footpoints close 
to the neutral line (red), and a set of over-arching loops (grey). Panel (c) shows the field lines close to the neutral line (red) together with the streamlines of the electric current density near the neutral line before the flare (yellow) and panel (d) shows the field lines (red) and current streamlines (yellow) after the flare. The coloured image in the background of each panel indicates the lower boundary values of $B_z$.\label{fig:f3}}
\end{figure}

To investigate the changes in the corona, we performed nonlinear force-free field (NLFFF) modeling of the magnetic field in the corona from the SHARP data, using the CFIT code~\citep{2007SoPh..245..251W} with the self-consistency procedure~\citep{2009ApJ...700L..88W,2011ApJ...728..112W}. Figure~\ref{fig:f3} shows coronal magnetic field solutions for the two times shown in Figure~\ref{fig:f1}, before and after the flare. Panel (a) shows field lines for the solution before the flare (17:34UT), and panel (b) shows field lines for the solution after the flare (18:58UT). The field lines in red, which are close to the neutral line, are more sheared in the post-flare solution. Panels (c) and (d) show the field lines close to the neutral line before and after the flare, respectively (red) as well as streamlines of the current density (yellow). The self-consistent solution is a close approximation to a force-free field, so the electric current density is everywhere parallel to the magnetic field. We note that results of NLFFF modeling for this region prior to the 22 June 2015 flare have also been presented by \citet{2017NatAs...1E..85W}. The results of our NLFFF calculations will be presented in more detail in a future publication.

\section{Model}\label{sec:model}

\subsection{Shear Alfv\'{e}n Wave}

\begin{figure}[t!h]
\begin{center}
\includegraphics[scale=0.5]{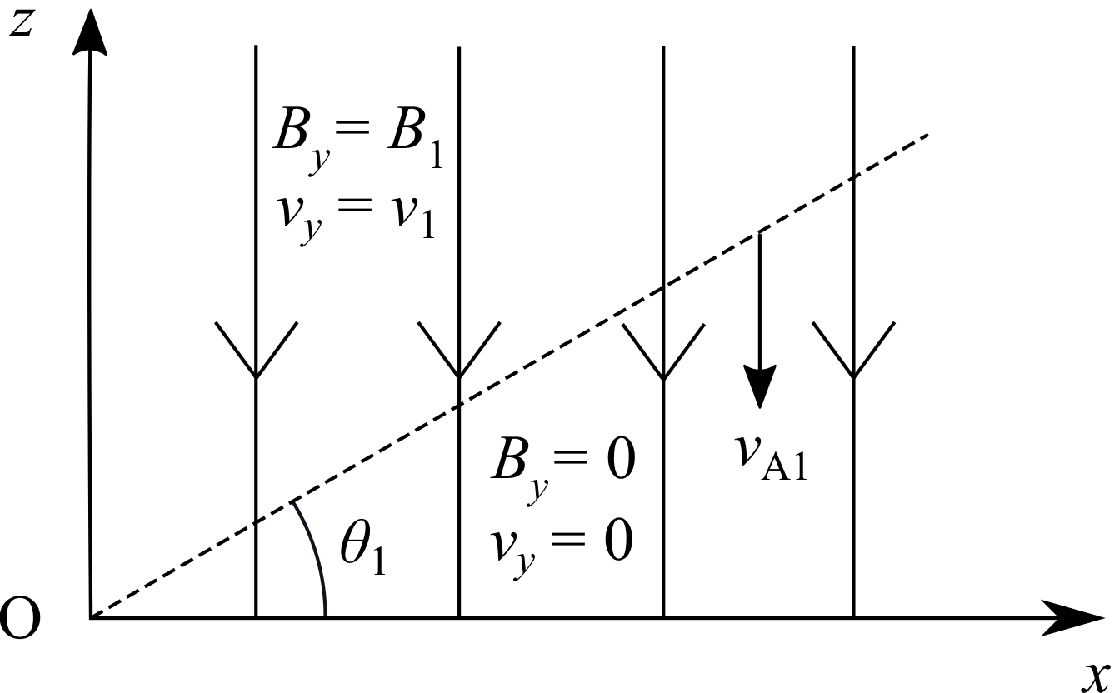}
\caption{The geometry of the model. A shear Alfv\'{e}n wave propagates down from the corona and impacts the photospheric plane ($z=0$). The Alfv\'{e}nic front is oriented at an angle $\theta_1$ as shown.\label{fig:f4}}
\end{center}
\end{figure}

We consider a simple 2-D model representing the response of the low solar atmosphere to a flare. Figure~\ref{fig:f4} illustrates the geometry of the model. The $z$-axis is the direction of the local vertical. The $x$-$y$ plane represents the photosphere.  The field lines shown indicate the local vertical component of the magnetic field. Before the flare, the field is assumed to be locally purely vertical, so $B_y=0$. The flare is assumed to introduce a shear component ($B_y\neq 0$), consistent with the discussion in Section~\ref{sec:analysis}. The shear propagates down from above, behind an Alfv\'{e}nic front, moving in the $-z$-direction with the local Alfv\'{en} speed $v_{\rm A1}$. The observations of differential sunspot rotation in AR 12371 presented by \citet{2016NatCo...713104L} showed clearly that the photosphere was set into motion behind the moving flare ribbon. To reproduce this aspect of the obervations we assume that the Alfv\'{e}nic front is at an angle $\theta_1$ to the $x$ axis, as shown. The $y$-axis (directed into the page) is  the location of the intersection of the front and the photosphere at the instant shown. The point of intersection moves to the right with time, representing the motion of the flare ribbon. This initial field configuration might be initiated by a process of reconnection which proceeds first for field lines with foot points close to the neutral line, and later for field lines with foot points further away from the neutral line.

To represent this model we assume a magnetic field of the form
\begin{equation}\label{eq:bform}
\bm{B}(x,z,t)=[0,B_y(x,z,t),-B_0],
\end{equation}
where $B_0$ is the magnitude of the constant vertical field, and $B_y(x,z,t)$ is the shear 
component. We assume also the form for the fluid velocity
\begin{equation}\label{eq:vform}
\bm{v}(x,z,t)=[0,v_y(x,z,t),0]
\end{equation}
where $v_y(x,z,t)$ is the flow associated with the shear.

With the assumed forms~(\ref{eq:bform}) and~(\ref{eq:vform}), the $y$-component of
the MHD equation of motion is
\begin{equation}\label{eq:mhd-motion}
\rho_1\frac{\partial v_y}{\partial t}=-\frac{B_0}{\mu_0}\frac{\partial B_y}{\partial z}
\end{equation}
where $\rho_1$ is the (assumed constant) coronal plasma density, and the ideal MHD induction equation is
\begin{equation}\label{eq:mhd-induction}
\frac{\partial B_y}{\partial t}=-B_0\frac{\partial v_y}{\partial z}.
\end{equation}
Equations~(\ref{eq:mhd-motion}) and~(\ref{eq:mhd-induction}) imply that both $B_y$ and
$v_y$ satisfy 1-D wave equations, e.g.\
\begin{equation}\label{eq:wave-equation}
\frac{\partial^2 B_y}{\partial t^2}=v_{\rm A1}^2\frac{\partial^2B_y}{\partial z^2},
\end{equation}
for $B_y$, where $v_{\rm A1}=B_0/\sqrt{\mu_0\rho_1}$.

Equation~(\ref{eq:wave-equation}) has the solution $B_y=f(z+v_{\rm A1}t)$ for any function $f$. This is the downward-propagating d'Alembert solution. The model shown in Figure~\ref{fig:f4} is reproduced with the specific choice:
\begin{equation}\label{eq:by-sol}
B_y(x,z,t)=B_1\theta (z+v_{\rm A1}t-\tan\theta_1x),
\end{equation}
where $\theta$ is the step function, $B_1$ is the shear component of the field, and time $t=0$ corresponds to the instant shown in Figure~4.
The corresponding velocity follows from Equations~(\ref{eq:mhd-motion}) 
or~(\ref{eq:mhd-induction}):
\begin{equation}
v_y(x,z,t)=-v_{\rm A1}\frac{B_1}{B_0}\theta (z+v_{\rm A1}t-\tan\theta_1 x).
\end{equation}
The shear Alfv\'{e}n wave introduces a velocity shear 
\begin{equation}\label{eq:v1}
v_1=-v_{\rm A1}\frac{B_1}{B_0}. 
\end{equation}
This is the Wal\'{e}n relationship for a shear Alfv\'{e}n wave propagating in the same direction as the background field.

The front introduces a horizontal component in the magnetic field and sets the fluid into motion. The power to do this is provided by a Poynting flux behind the front, directed downwards. To see this note that the electric field behind the front is $\bm{E}=-\bm{v}\times\bm{B} =v_1B_0\hat{\bf x} =-v_{\rm A1}B_1\hat{\bf x}$. The Poynting flux is 
\begin{equation}
    \bm{P}_{\rm P1}=\frac{1}{\mu_0}\bm{E}\times\bm{B}=-\frac{v_{\rm A1}B_1^2}{\mu_0}\hat{\bf z}.
\end{equation}
The increase in energy per unit time and per unit area in the $x-y$ plane due to the introduction of the shear component of the magnetic field at the front is given by
\begin{equation}\label{eq:PB1}
    P_{B1}=\frac{1}{2\mu_0}B_1^2v_{\rm A1}.
\end{equation}
Similarly the power per unit area associated with the kinetic energy introduced at the front is given by
\begin{equation}\label{eq:PK11}
    P_{K1}=\frac{1}{2}\rho_1v_1^2v_{\rm A1},
\end{equation}
and using $v_1=-v_{\rm A1}B_1/B_0$ we have
\begin{equation}\label{eq:PK12}
    P_{K1}=\frac{1}{2\mu_0}B_1^2v_{\rm A1}.
\end{equation}
Equations~(\ref{eq:PB1})--(\ref{eq:PK12}) are independent of $\theta_1$ because in a time $\Delta t$ the front crosses an area in the $x-z$ plane, per unit length in $x$, which depends only on $v_{\rm A1}\Delta t$.

Hence we have $P_{B1}=P_{K1}$ and $|P_{\rm P1}|=P_{B1}+P_{K1}$. The Poynting flux accounts for the increase in magnetic and kinetic energy at the front, and there is the usual equipartition between magnetic and kinetic energy in a shear Alfv\'{e}n wave.

\subsection{Photospheric Response}

To model the photospheric response, we represent the sub-photosphere ($z<0$) as a region with (uniform) plasma density $\rho_2$ and Alfv\'{e}n speed $v_{\rm A2}$. We again treat the plasma as being ideal, so that Equations~(\ref{eq:bform})--(\ref{eq:wave-equation}) describe the magnetic field and velocity shear in the sub-photosphere, with $\rho_1$ and $v_{\rm A1}$ replaced by $\rho_2$ and $v_{\rm A2}$.

\begin{figure}[t!h]
\begin{center}
\includegraphics[scale=0.5]{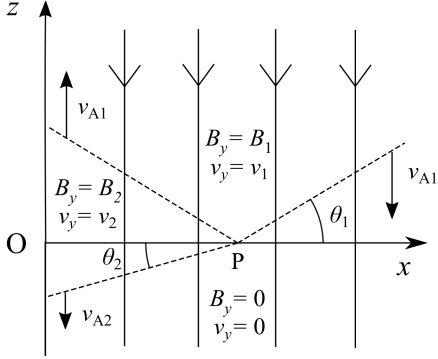}
\caption{The reflection and transmission of the Alfv\'{e}nic front at the photosphere.\label{fig:f5}}
\end{center}
\end{figure}

When the front reaches the photosphere it is partly reflected and partly transmitted. Figure~\ref{fig:f5} illustrates the situation at a time $t>0$. The downward propagating front reaches the photosphere at the point P, located at $x_{\rm P}=v_{\rm A1}t/\tan\theta_1$. For $x<x_{\rm P}$ there are reflected and transmitted fronts in the corona and sub-photosphere respectively. The front moves more slowly in the sub-photosphere, so the transmitted front is inclined at an $\theta_2$ to the $x$-axis, where $v_{\rm A1}/\tan\theta_1=v_{\rm A2}/\tan\theta_2$.

We assume the field and velocity shear components between the fronts in the region $x< x_{\rm P}$ are $B_y=B_2$ and $v_y=v_2$ respectively. The horizontal component of the magnetic field must be continuous across the photospheric boundary because there cannot be a static current in the boundary. The horizontal velocity component in the model must also be continuous across the boundary.

The shear component of the field in the corona after reflection (the region $x<x_{\rm P}$ and $z\geq 0$) is given by
\begin{equation}
B_y(x,z,t)=B_2+(B_1-B_2)\theta (z-v_{\rm A1}t+\tan\theta_1x).
\end{equation}
Applying Equation~(\ref{eq:mhd-motion}) gives
\begin{equation}\label{eq:v-reflected-front}
v_y(x,z,t)=v_{\rm A1}\frac{(B_1-B_2)}{B_0}\theta (z-v_{\rm A1}t+\tan\theta_1 x)+v_2.
\end{equation}
Ahead of the reflected front we require $v_y=v_1=-v_{\rm A1}B_1/B_0$, so Equation~(\ref{eq:v-reflected-front}) implies
\begin{equation}\label{eq:v21}
v_2=-v_{\rm A1}\frac{(2B_1-B_2)}{B_0}.
\end{equation}

The shear component of the field in the sub-photosphere after transmission (the region $x<x_{\rm P}$ and $z<0$) is given by
\begin{equation}
B_y(x,z,t)=B_2\theta (z+v_{\rm A2}t-\tan\theta_2x).
\end{equation}
Applying Equation~(\ref{eq:mhd-motion}) gives
\begin{equation}\label{eq:v-transmitted-front}
v_y(x,z,t)=-v_{\rm A2}\frac{B_2}{B_0}\theta (z+v_{\rm A2}t-\tan\theta_2x).
\end{equation}
Equations~(\ref{eq:v-reflected-front}) and~(\ref{eq:v-transmitted-front}) must match at $z=0$, which implies
\begin{equation}\label{eq:v22}
v_2=-v_{\rm A2}\frac{B_2}{B_0}.
\end{equation}

Equations~(\ref{eq:v21}) and~(\ref{eq:v22}) imply
\begin{equation}\label{eq:B2-v2}
    B_2=\frac{2v_{\rm A1}}{v_{\rm A1}+v_{\rm A2}}B_1 \quad \mathrm{and}\quad v_2=\frac{2v_{\rm A2}}{v_{\rm A1}+v_{\rm A2}}v_1.
\end{equation}
These relations define the photospheric response in the model. In the limit of an infinitely dense photosphere ($v_{\rm A2}\rightarrow 0$) we have $B_2\rightarrow 2B_1$ and $v_2\rightarrow 0$: the shear Alfv\'{e}n wave is completely reflected. Otherwise, the wave is partly transmitted and partly reflected, with $B_2>B_1$ and $v_2<v_1$.

In the model the shear Alfv\'{e}n wave is propagating in the direction of the field. Equations~(\ref{eq:v1}) and~(\ref{eq:v22}) relate the velocity and field amplitudes in the wave. We have $v_i=-v_{\rm A\it i}B_i/B_0$, with $i=1,2$. If a shear Alfv\'{e}n wave is propagating in the opposite direction to the field the relationships are $v_i=v_{\rm A\it i}B_i/B_0$ with $i=1,2$. This situation applies if we consider an Alfv\'{e}nic front also propagating down from the corona to the photosphere on the opposite side of the neutral line. On the negative polarity side of the neutral line, the shear components of the field and the fluid flow behind the front have a different sign, and on the positive polarity side they have the same sign. This may also be understood in terms of the Poynting flux. On both sides of the neutral line the Poynting flux is directed downwards, to provide the energy for the changes in the field and flow. For the given geometry the $z$-component of the Poynting vector is $P_{\rm P}=-v_yB_yB_z/\mu_0$, so if $B_z$ is positive, a downwards (upwards) directed Poynting flux implies $v_yB_y>0$ ($v_yB_y<0$).

\subsection{Currents in the Model}

The electric current density for the model geometry is
\begin{equation}\label{eq:current-density}
\bm{J}(x,z,t)=\frac{1}{\mu_0}
  \left(-\frac{\partial B_y}{\partial z},0,\frac{\partial B_y}{\partial x}\right).
\end{equation}
The model includes surface currents in the Alfv\'{e}nic fronts, and, if ${\rm d}B_1/{\rm d}x$ and ${\rm d}B_2/{\rm d}x$ are non-zero, a vertical current density behind and between the fronts. If we consider the currents at the photosphere ($z=0$), then, applying Equation~(\ref{eq:current-density}), we find a surface current at the location of the front:
\begin{equation}\label{eq:current-front}
K^{\rm F}_z=-\frac{1}{\mu_0}B_2(x=x_{\rm P})
\end{equation}
and a vertical current density behind the front:
\begin{equation}\label{eq:jz-behind}
J_z^{\rm BF}=\frac{1}{\mu_0}\frac{{\rm d}B_2}{{\rm d}x}\theta (v_{\rm A1}t-\tan\theta_1x).
\end{equation}

Equation~(\ref{eq:jz-behind}) represents the current density which appears close to the neutral line after the flare, as seen in Figure~\ref{fig:f1}. The observed current density has an average value $J^{\rm BF}_z\approx-25$\,mA/m$^2$. This implies a gradient in the shear  ${\rm d}B_2/{\rm d}x=\mu_0J^{\rm BF}_z\approx -3.1\times 10^{-8}$\,T/m. The current density which appears has a lateral extent of order $L\approx 5^{\arcsec}\approx 3.6\times 10^6$\,m, so over this length scale the field gradient implies a change in the field of about $({\rm d}B_2/{\rm d}x)L\approx 1100$\,gauss ($0.11\,$T), which is consistent with the changes seen in Figure~\ref{fig:f2}. Hence the simple model gives a consistent description of the observed changes in the field and the associated currents.


\section{Discussion}\label{sec:discussion}

The model presented here is highly simplified, but it is able to represent observed features of the photospheric response to a flare. In the model, an increased magnetic shear is introduced along the neutral line due to a downward propagating shear Alfv\'{e}n wave, with the change occurring at the photosphere behind a moving front. The increase in magnetic shear coincides with the appearance of velocity shear, which is oppositely directed on either side of the neutral line, and also with the appearance of a vertical current density, in the case that the magnetic shear varies with distance from the neutral line.

The model is 2-D, but the actual geometry is of course more complex. Figure~\ref{fig:f2} shows that the rotating sunspot in AR 12371 involves changes in the horizontal field at the photosphere which curl around the spot. However, the simple model provides a basis for understanding the real process.

The wave equation is lossless, so the changes introduced by the shear Alfv\'{e}n wave are reversible. The constant values $B_1$ and $v_1$ of the shear components of the magnetic field and flow in the initial downward-propagating wave can be thought of as being maintained by boundary conditions above, at an upper boundary at $z=L$, say. The model has $B_y(x,z=L,t)=B_1$ for $t>0$, until the time at which the reflected front returns to $z=L$. The boundary conditions can be thought of as a continual driving at $z=L$. If the driving switches off [$B_y(x,z=L,t)=0$] a new front is launched which removes the shear components below. This is a somewhat artificial aspect of the model, but the model is intended only to show how a shear Alfv\'{e}n wave can introduce sudden sub-photospheric changes matching the flare observations. Accurate modeling of the longer-time evolution of the system is expected to require a more realistic geometry, and explicit prescription of the boundary and initial conditions. Also, a more realistic model should include loss, and this can produce irreversible change.

Based on magnetograms constructed in the near-infrared, \citet{Xu-etal-2018} reported transient horizontal field changes in the 22 June 2015 flare, as well as the permanent changes discussed here. The transient changes occurred at certain locations close to the neutral line, as the flare ribbons passed. The authors argued that the changes may be due to torsional Alfv\'{e}n waves generated by reconnection propagating down from above. This picture is similar to the model developed here. In our model, transient changes can be produced by impulsive, rather than continual, driving from above, so in principle the model can account also for changes of this kind.

The relationship between the change in the magnetic field and the velocity at the photosphere in the model is $v_2/v_{\rm A2}=-B_2/B_0$. Assuming a photospheric mass density $\rho_2=5\times 10^{-4}$ kg/m$^3$ and a vertical field $B_0=1000$ gauss ($0.1\,$T) gives an Alfv\'{e}n speed $v_{\rm A2}=4\times 10^3$ m/s. The flow velocities for AR 12371 obtained by tracking \citep{2016NatCo...713104L,2018ApJ...853..143W} are $v_2\approx 0.1-1\times 10^3$ m/s. The shear wave relationship then implies $|B_2|\approx 25-250$ gauss, which is consistent with the observations (see Figure~\ref{fig:f2}). Hence the shear-wave model accounts for the relative sizes of the observed changes in the photospheric magnetic field and plasma motion.

The shear Alfv\'{e}n wave has a downwards-directed Poynting flux $P_{\rm P1}=v_{\rm A1}B_1^2/\mu_0$ in the corona. We can estimate the total implied energy deposition during the flare from the observations for AR 12371. Figure~\ref{fig:f2} implies a change in the horizontal field $B_2\approx 200$ gauss $=2\times 10^{-2}$ T over an area around the neutral line $A\approx 0.5\times 1$ deg$^2\approx  10^{15}$ m$^2$. Assuming $v_{\rm A1}\gg v_{\rm A2}$ in Equations~(\ref{eq:B2-v2}) we have $B_1\approx 0.5B_2=\times 10^{-2}$ T. Assuming a coronal Alfv\'{e}n speed $v_{\rm A1}\approx 10^6$ m/s the implied power is $P_{\rm P1}A\approx 8\times 10^{22}$ W. Over the time scale $T=60$ s of the flare this implies deposition of a total energy $P_{\rm P1}AT\approx 5\times 10^{24}$ J, which is comparable to the total energy in a large flare. \citet{2016NatCo...713104L} used flows obtained by tracking to calculate the Poynting flux at the photosphere in AR 12371 over a more limited area, and identified a net downwards flux during the flare with total energy $1.6\times 10^{23}$ J.

The changes observed at the photosphere occur behind the moving flare ribbons, which coincide with the location of hard X-ray production at the photosphere~\citep{2017ApJ...842L..18J}. An intriguing possibility is that the changes in the low atmosphere play a role in particle acceleration. In the model the Alfv\'{e}nic front carries a surface current. The current density in the front implied by Equation~(\ref{eq:current-front}):
\begin{equation}
J_z^{\rm F}=\frac{K_z^{\rm F}}{\ell/\sin\theta_1}
\end{equation}
may be large if the thickness $\ell$ of the front is small. In the low atmosphere the gas is partially ionised, and has a conductivity much less than the fully-ionised corona. This allows the possibility of a significant field-aligned electric field. The classical parallel conductivity is dominated by the contribution from
electron-neutral collisions~\citep{Huba2013}:
\begin{equation}
\sigma_{\parallel {\rm e}}=\frac{n_{\rm e}e^2}{m_{\rm e}\nu_{\rm ne}},
\end{equation}
where $n_{\rm e}$ is the electron number density, and $\nu_{{\rm ne}}$ is the electron-neutral collision frequency. The field-aligned electric field implied by this conductivity is 
\begin{equation}
\begin{split}
E_{\parallel}^{\rm F}(t)&=J^{\rm F}_z(t)/\sigma_{\parallel {\rm e}}\\
  &=-\sin\theta_1\frac{B_2(v_{\rm A1}t/\tan\theta_1)}{\mu_0\ell}\frac{m_{\rm e}\nu_{\rm ne}}{ne^2}.
\end{split}
\end{equation}
A critical electric field for electron runaway may be estimated by the balance between
the electric force and the drag force due to collisions of a thermal electron with
neutrals:
\begin{equation}
E_{\rm c}=\frac{m_{\rm e}v_{\rm e}\nu_{\rm n_{\rm e}e}}{e}.
\end{equation}
The ratio of the field-aligned electric field due to the current density in the front and 
the critical field is
\begin{equation}
\begin{split}
\frac{\left|E_{\parallel}^{\rm F}(t)\right|}{E_{\rm c}}&=
  \sin\theta_1
  \frac{\left|B_2(v_{\rm A}t/m)\right|}{\mu_0\ell}\frac{1}{nev_{\rm e}}\\
  &=\frac{\left|J_z^{\rm F}(t)\right|}{n_{\rm e}ev_{\rm e}}.
\end{split}
\end{equation}
We can estimate this ratio using chromospheric values for the atmospheric
parameters:
\begin{equation}
\frac{\left|E_{\parallel}^{\rm F}\right|}{E_{\rm c}}
  \approx 1.3 \left(\frac{\left|B_2^{\rm F}\right|}{10^{-2}\,{\rm T}}\right)
    \left(\frac{10\,{\rm m}}{\ell}\right)\left(\frac{10^4\,{\rm K}}{T_{\rm e}}\right)^{1/2}.
\end{equation}
This suggests that the electric field may exceed the critical field if the front has a sufficiently narrow width ($\approx 10$\,m). This estimate relies on the use of the classical conductivity/collision frequency: it is also possible that an anomalous resistivity associated with an effective collision frequency due to turbulence or microphysical structures is relevant, in which case the width could be much greater~\citep[e.g.][]{2012ApJ...749..166H}.

\begin{figure}[t!h]
\begin{center}
\includegraphics[scale=0.5]{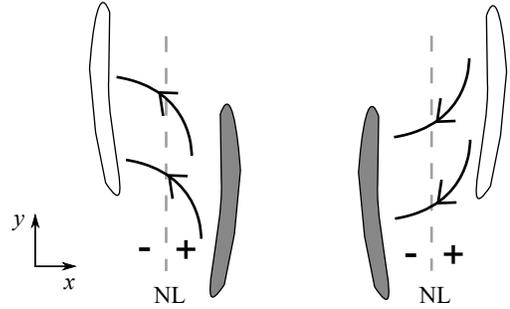}
\caption{Two possible flare configurations. Two sheared field lines are shown crossing the magnetic neutral line (NL). Flare ribbons are shown on either side of the neutral line. The model predicts an asymmetry in hard X-ray production, with more emission at the shaded flare ribbon. \label{fig:f6}}
\end{center}
\end{figure}

If the shear Alfv\'{e}n wave produces electron acceleration in the low atmosphere, then energy is transported Alfv\'{e}nically from the corona, and then locally dissipated. This idea has been proposed before~\citep[e.g.][]{2008ApJ...675.1645F,2013SoPh..288..223M}. An attractive feature of this picture is that it avoids the number problem posed by acceleration in the corona. A prediction of the present model is that the field aligned electric field component is directed downwards (towards the photosphere) on one side of the neutral line, and upwards (away from the photosphere) on the other side. Hard X-ray production by bremsstrahlung will then occur predominantly at the foot point with the field directed upwards, which implies an asymmetry in hard X-ray production at the two flare ribbons on either side of the neutral line. For the configuration shown in Figures~\ref{fig:f4} and \ref{fig:f5}, with the shear field $B_1$ in the $y$-direction on the negative polarity side of the neutral line, the electric field is directed downwards in the front [Equation (25)]. Hence the HXR production is expected to occur predominantly on the other (positive) polarity side. However, if the shear field is instead in the negative $y$-direction on the negative polarity side, this is expected to be the side where most hard X-rays are produced. Figure~\ref{fig:f6} illustrates the expected asymmetry. The left-hand panel corresponds to the configuration in Figures~\ref{fig:f4} and \ref{fig:f5}.
The situation is analogous to auroral particle acceleration. It is well established in the Earth's magnetosphere that acceleration of electrons by $E_\parallel$ occurs only on the upward current path, and not on the neighbouring downward current path.

In the present model we assume that a shear component of the field is introduced close to the neutral line by a downward-propagating shear Alfv\'{e}n wave. There are specific models for eruptions which involve the appearance of sheared fields along the neutral line after a flare/eruption. In the ``tether-cutting'' model, reconnection of field lines close to the neutral line leads to the formation of low-lying, sheared loops~\citep{2001ApJ...552..833M}. In the ``magnetic implosion'' picture~\citep{2000ApJ...531L..75H} a reduction in magnetic pressure due to loss of magnetic energy is assumed to lead to a more compact magnetic structure over the neutral line, with more horizontal fields. These models attempt to explain the origin of the increased shear in the corona. We have not tried to do that, but have instead focused on how the shear is transmitted from the corona to the sub-photosphere.


\section{Conclusions}\label{sec:conclusions}

We present a 2-D model which explains the sudden appearance of magnetic and velocity shear at the photosphere during a flare in terms of a downwards-directed large amplitude Alfv\'{e}n shear wave impacting the photosphere on either side of the neutral line. The shear Alfv\'{e}n wave is assumed to be produced by magnetic field reconnection in the flare. Although the wave propagates vertically downwards, the front is assumed to be inclined to the photosphere, so that the front arrives first at locations closer to the neutral line. This is intended to reproduce the observations of a sudden photospheric response to a flare behind spreading flare ribbons~\citep{2016NatCo...713104L}.

The model front is transmitted and reflected at the photosphere, and the transmitted wave introduces a horizontal component in the magnetic field, and a horizontal flow, beneath the photosphere. In principle this can account for the surprising observations of sudden motion of the photosphere in response to a flare~\citep{2016NatPh..12..998A}. The model predicts that the shear introduced by the wave in the photospheric magnetic field has the same sign on either side of the neutral line, whereas the velocity shear has the opposite sign. Also, the total energy deposited in the photosphere by the shear Alfv\'{e}n wave is comparable to the flare energy. We speculate that the sudden changes in the magnetic field in the low atmosphere are associated with particle acceleration in the flare.

The model is highly simplified, but in principle it can account for a range of effects due to a flare. It remains to work out the details, and to develop more detailed models.

\acknowledgments
We thank the SDO/HMI team for the HMI data. SDO is a mission in NASA's Living With a Star (LWS) Program. We thank also Nastaran Farhang for comments on the manuscript. This work was funded in part by an Australian Research Council Discovery Project (DP160102932).

\end{document}